\documentclass[twocolumn,final]{IEEEtran}
\usepackage{cite}
\usepackage[cmex10]{amsmath}
\usepackage{amsthm}
\usepackage{amssymb}
\usepackage{balance}
\usepackage[final]{graphicx}
\usepackage{color}
\usepackage{epsfig}
\usepackage{epstopdf}
\usepackage{tikz}
\usetikzlibrary{positioning}
\usepackage[ruled,vlined,linesnumbered]{algorithm2e}

\newcommand{\bs}[1]{\boldsymbol{#1}}
\newcommand{\mc}[1]{\mathcal{#1}}

\newcommand{\mb}[1]{\mathbf{#1}}
\newcommand{\mr}[1]{\mathrm{#1}}
\newcommand{\tr}{\mathrm{Tr}}

\begin{document}

\title{\LARGE Neural Network-Optimized Channel Estimator and Training Signal Design for MIMO Systems with Few-Bit ADCs}
\author{Duy H. N. Nguyen, \emph{IEEE, Senior Member}
	\thanks{Duy H. N. Nguyen is with the Department of Electrical and Computer Engineering, San Diego State University, San Diego, CA, USA 92182. E-mail: duy.nguyen@sdsu.edu.}}

\maketitle

\begin{abstract}
This paper is concerned with channel estimation in MIMO systems with few-bit ADCs. In these systems, a linear minimum mean-squared error (MMSE) channel estimator obtained in closed-form is not an optimal solution. We first consider a deep neural network (DNN) and train it as a nonlinear MMSE channel estimator for few-bit MIMO systems. We then present a first attempt to use DNN in optimizing the training signal and the MMSE channel estimator concurrently. Specifically, we propose an autoencoder with a specialized first layer, whose weights embed the training signal matrix. Consequently, the trained autoencoder prompts a new training signal designed specifically for the MIMO channel model under consideration.
\end{abstract}

\vspace{-0.3cm}
\section{Introduction}
In multiple-input multiple-output (MIMO) systems, accurate channel state information (CSI), acquired from channel estimation, is crucial to unleash the gain of MIMO communications \cite{Tse-BOOK}. In massive MIMO systems, the availability of CSI enables the base-station (BS) to reduce the effects of noise and interference, and thus lead to improvements in spectral efficiency and energy efficiency \cite{Rusek-SPMag-2013}. CSI estimation at a receiver is typically performed during a training phase, when a known training sequence is sent from a transmitter. We note that the deep literature of MIMO signal processing has effectively resolved the channel estimation problem with rigorous performance analysis, especially in linear MIMO systems. 

Recent research in massive MIMO advocated for the use of low-resolution, i.e., $1$--$3$ bits, analog-to-digital converters (ADCs) to reduce the power consumption at wireless transceivers \cite{Li-Swindlehurst-TSP17,Mo-Heath-TSP18}. However, the severe nonlinear distortion induced by few-bit ADCs can make the channel estimation task very challenging \cite{Mo-Heath-TSP18}. A common approach to tackle this problem was to first linearize the coarsely quantized signal by the Bussgang decomposition \cite{Bussgang-1952}. A closed-form Bussgang-based linear minimum mean-squared error (BLMMSE) channel estimator was then proposed \cite{Mollen-TWC17,Li-Swindlehurst-TSP17}. However, a BLMMSE channel estimator is \emph{not} optimal since the quantized observation is not Gaussian. Generalized approximate message passing (GAMP) is another approach for channel estimation with low-bit observations \cite{Mo-Heath-TSP18}.

The nonlinearity in coarse quantization can also be well captured by deep neural networks (DNN). Recent work in \cite{Gao-Li-CLetters-2019} proposed deep learning based channel estimation for massive MIMO systems with mixed-resolution ADCs. Another work in \cite{Takeda-DNN-ADC-2019} proposed a DNN-based channel estimator for one-bit non-ideal ADCs with threshold hysteresis. DNN was also considered for estimating the effective channel in massive MIMO systems under hardware nonlinearity in a recent paper \cite{Bjornson-OJ-2020}. All these papers showed promising results by DNN-based channel estimation over existing analytical methods, such as BLMMSE and GAMP. Interestingly, regression with DNN can be interpreted as a nonlinear MMSE estimator \cite{EE-278}, which thus facilitates a data-driven approach for channel estimation. However, to the best of our knowledge, none of existing work considered optimizing the training signal for MIMO systems with few-bit ADCs. 

In this work, we first consider a feed-forward DNN regressor as a nonlinear MMSE estimator for MIMO channels with few-bit observations. More importantly, different from existing work in using DNN for channel estimation, we propose a DNN autoencoder to jointly optimize the training sequence and the channel estimator. The proposed autoencoder includes a specialized first layer whose weights embed the training signal matrix. Once being trained, the autoencoder prompts a DNN-optimized training signal design using the data generated by the channel model under consideration. We then present numerical results for two channel models: i.) Rayleigh fading with independent and identically distributed (i.i.d.) complex Gaussian channel coefficients and ii.) light-of-sight (LoS) channel coefficients. Numerical results show superior performance in terms of mean-squared error (MSE) by the DNN-optimized training signal design over discrete Fourier transform (DFT)-based orthogonal training sequences.
 
\vspace{-0.25cm}
\section{System Model}
We consider the training phase over $\tau$ time slots from $K$ transmit antennas to an $M$-antenna base-station. The unquantized system model can be formulated as
\vspace{-0.15cm}
\begin{eqnarray}\label{system-model}
\mb{Y} = \sqrt{\rho}\,\mb{H}\mb{\Phi}^T  + \mb{N} 
\end{eqnarray}
where $\mb{\Phi} \in \mathbb{C}^{\tau\times K}$ is the training pilot transmitted from the $K$ antennas, $\rho$ is a power scaling factor at the transmitter, $\mb{Y} \in \mathbb{C}^{M\times \tau}$ is the unquantized received signal, $\mb{N}$ is the additive noise, and $\mb{H}  \in \mathbb{C}^{M\times K}$ is the channel matrix to be estimated. This system model is applicable for training from $K$ single-antenna users or from a single $K$-antenna user. If the pilot sequences $[\mb{\Phi}]_i, i=1,\ldots,K$, where $[\mb{\Phi}]_i$ is the $i$th column of $\mb{\Phi}$,  are drawn from $K$ columns of an $\tau\times\tau$ DFT matrix, they are orthogonal with each other, i.e., $\mb{\Phi}^H\mb{\Phi} = \tau \mb{I}_K$. This work, however, is not limited to the case of orthogonal pilot sequences. Instead, we set power constraints on the design of $\mb{\Phi}$. For a multiuser system, a per-user power constraint is assumed such that $[\mb{\Phi}]_i^H[\mb{\Phi}]_i \leq \tau$.  For a single-user system, either a per-antenna power constraint $[\mb{\Phi}]_i^H[\mb{\Phi}]_i \leq \tau$ or a sum-power constraint $\tr\{\mb{\Phi}^H\mb{\Phi}\} \leq \tau K$ is assumed.  
For ease of representation, we vectorize the system model \eqref{system-model} into
\vspace{-0.15cm}
\begin{eqnarray}\label{vector-model}
\mb{y} = \mr{vec}(\mb{Y}) = \bar{\mb{\Phi}} \mb{h}+ \mb{n} 
\end{eqnarray}
where $\bar{\mb{\Phi}} = \sqrt{\rho} \mb{\Phi}\otimes \mb{I}_M$  with $\otimes$ being the Kronecker product, $\mb{h} = \mr{vec}(\mb{H})$, and $\mb{n} = \mr{vec}(\mb{N})$. We assume that the channel vector $\mb{h}$ is comprised of random variables with zero mean and covariance matrix of $\mb{C}_{\mb{h}}$
and the noise vector $\mb{n}$ is $\mc{CN}(\mb{0},\mb{C}_{\mb{n}})$.  We then define the SNR as $\frac{\rho\; \mathbb{E}\{\|\mb{h}\|^2\}}{\mathbb{E}\{\|\mb{n}\|^2\}} = \frac{\rho\tau\,\tr\{\mb{C}_{\mb{h}}\}}{K\,\tr\{\mb{C}_{\mb{n}}\}}$. 
If $\mb{y}$ is quantized by a $b$-bit uniform  quantizer $\mc{Q}_b(\cdot)$, we obtain the quantized signal $\mb{r} = \mc{Q}_b(\mb{y})$. The focus of this work is to find a channel estimator $\mb{\hat{\mb{h}}}$ to minimize the MSE $\mathbb{E}\big\{\|\mb{\hat{\mb{h}}}-\mb{h}\|^2\big\}$, given the observation $\mb{r}$.
\vspace{-0.25cm}
\section{MMSE Channel Estimator for MIMO Systems}
\subsection{Channel Estimation and Training Signal Design in Unquantized MIMO Systems}
With unquantized signal $\mb{y}$, an MMSE estimator can be uniquely defined as $\hat{\mb{h}}_{\texttt{MMSE}}= \mathbb{E}\,[\mb{h}|\mb{y}]$. It is often difficult to find an optimal MMSE estimator in closed-form. One alternative approach is to find an optimal linear MMSE estimator, which is given by $\hat{\mb{h}}_{\texttt{LMMSE}} = \mb{C}_{{\mb{h}}\mb{y}}\mb{C}_{\mb{y}}^{-1}\mb{y}$. Here, $\mb{C}_{\mb{hy}} = \mathbb{E}[\mb{h}\mb{y}^H] = \bar{\mb{\Phi}}\mb{C}_{\mb{h}} \bar{\mb{\Phi}}^H + \mb{C}_{\mb{n}}$ is the cross-covariance matrix between $\mb{h}$  and $\mb{y}$, and  $\mb{C}_{\mb{y}} = \mathbb{E}[\mb{y}\mb{y}^H] = \mb{C}_{ \mb{h}}\bar{\mb{\Phi}}^H$ is the covariance matrix of $\mb{y}$.  The linear MMSE estimator then can be written as \cite{Hassibi-TIT-2003}
\vspace{-0.15cm}
\begin{eqnarray}\label{LMMSE}
\hat{\mb{h}}_{\texttt{LMMSE}} &=& \mb{C}_{ \mb{h}}\bar{\mb{\Phi}}^H\left(\bar{\mb{\Phi}}\mb{C}_{\mb{h}} \bar{\mb{\Phi}}^H + \mb{C}_{\mb{n}}\right)^{-1}\mb{y}\nonumber \\
&=& \left(\bar{\mb{\Phi}}^H\mb{C}_{\mb{n}}^{-1}\bar{\mb{\Phi}} + \mb{C}_{\mb{h}}^{-1}\right)^{-1}\bar{\mb{\Phi}}^H\mb{C}_{\mb{n}}^{-1}\mb{y}.
\end{eqnarray}

The covariance matrix of the estimation error vector $\bs{\varepsilon} = \mb{h} - \hat{\mb{h}}_{\texttt{LMMSE}}$  can be found to be
\vspace{-0.15cm}
\begin{eqnarray}
\mb{C}_{\bs{\varepsilon}} = \mb{C}_{\mb{h}} -  \mb{C}_{\mb{h}}\bar{\mb{\Phi}}^H\mb{C}_{\mb{y}}^{-1}\bar{\mb{\Phi}}^H\mb{C}_{\mb{h}} =  \left(\bar{\mb{\Phi}}^H\mb{C}_{\mb{n}}^{-1}\bar{\mb{\Phi}} + \mb{C}_{\mb{h}}^{-1}\right)^{-1}.
\end{eqnarray}

It is worth mentioning that the linear MMSE estimator is the \emph{optimal} MMSE estimator if $\mb{y}$ and $\mb{h}$ are jointly Gaussian distributed \cite{Hassibi-TIT-2003}. This is the case when $\mb{h}$ and $\mb{n}$ are both Gaussian. Then, the task of optimizing the training matrix $\mb{\Phi}$ is to minimize the total MSE $\tr\{\mb{C}_{\bs{\varepsilon}}\}$, subject to power constraint(s) on $\mb{\Phi}$. When the channel vector and the noise vector are both i.i.d. Gaussian, e.g.,  $\mb{C}_{\mb{h}} = \mb{I}_{MK}$ and $\mb{C}_{\mb{n}} = N_0\mb{I}_{M\tau}$, the optimal training signal must have orthogonal columns \cite{Hassibi-TIT-2003}.  It thus suffices  to choose $\mb{\Phi}$ as $K$ columns of a $\tau\times \tau$ DFT matrix. 
However, optimizing the training matrix $\mb{\Phi}$ can be much more involved for non-i.i.d. channel and/or noise vectors, even if they are both Gaussian. Several papers proposed closed-form solutions to the training signal $\mb{\Phi}$ under the sum power constraint, such as \cite{Gershamn-TSP-2006} for correlated channels and white noises, \cite{Liu-TSP2007} for correlated channels and colored noises, and \cite{Bjornson-TSP-2010}  for Rician fading channels and nonzero-mean colored noises. 
\vspace{-0.25cm}
\subsection{Channel Estimation in Low-Bit MIMO Systems} \label{quantized-sect}
Suppose that the received signal $\mb{y}$ is quantized by a $b$-bit uniform quantizer $\mc{Q}_b(\cdot)$ providing the observation $\mb{r} = \mc{Q}_b(\mb{y})$, where the quantization is applied separately to the real and imaginary parts of $\mb{y}$. An MMSE estimator is then given by $\hat{\mb{h}}_{\texttt{MMSE}}^{\mc{Q}} = \mathbb{E}\{\mb{h}|\mb{r}\}$. Since the quantized vector $\mb{r}$ is not Gaussian, finding an optimal MMSE estimator can be challenging. To circumvent this difficulty, recent work on low-bit MIMO systems \cite{Li-Swindlehurst-TSP17,Mollen-TWC17} relied on linearizing the nonlinear quantization operator $\mc{Q}_b(\cdot)$ using the  Bussgang decomposition \cite{Bussgang-1952,Mezghani-ISIT12}. Assuming that the quantizer input $\mb{y}$ is Gaussian distributed, one can decompose $\mb{r}$ into a desired signal component of $\mb{y}$ and an uncorrelated distortion $\mb{e}$ \cite{Mezghani-ISIT12,Mo-TSP-2017} such that
\vspace{-0.15cm}
\begin{eqnarray}\label{Bussgang}
\mb{r}= (1-\eta_b) \mb{y} + \mb{e} = (1-\eta_b)\mb{\bar{\Phi}}\mb{h} + (1-\eta_b)\mb{n} + \mb{e}
\end{eqnarray}
where $\eta_b$ is a distortion factor. The value of $\eta_b$ and the step-size $\Delta_q$ of an optimal $b$-bit uniform quantizer with a unit-variance Gaussian input is given in Table \ref{Table-1}. We also include in the table an optimal ternary quantizer with $3$ quantizing levels $\{-1.224,0,1.224\}$. Note that the step-size $\Delta_b$ for determining the decision levels must be scaled by the standard deviation of the input source.


A BLMMSE estimator is given by $\hat{\mb{h}}^{\mc{Q}_b}_{\texttt{BLMMSE}} = \mb{C}_{\mb{hr}}\mb{C}_{\mb{r}}^{-1}\mb{r}$, where $\mb{C}_{\mb{hr}} = \mathbb{E}[\mb{h}\mb{r}^H]$ and $\mb{C}_{\mb{r}} = \mathbb{E}[\mb{r}\mb{r}^H]$.
Since $\mb{h}$ and $\mb{e}$ are uncorrelated \cite{Li-Swindlehurst-TSP17}, one has $\mb{C}_{\mb{hr}} = (1-\eta_b)\mb{C}_{\mb{h}}\bar{\mb{\Phi}}^H$.  The covariance matrix of the channel estimation error  $\bs{\varepsilon} = \mb{h} - \hat{\mb{h}}^{\mc{Q}}_{\texttt{BLMMSE}}$ is then given by
$\mb{C}_{\bs{\varepsilon}} = \mb{C}_{ \mb{h}} - (1-\eta_b)^2 \mb{C}_{ \mb{h}}\bar{\mb{\Phi}}^H\mb{C}_{\mb{r}}^{-1}\bar{\mb{\Phi}}\mb{C}_{ \mb{h}}.$

For the case of symmetric $1$-bit quantizing, the covariance matrix $\mb{C}_{\mb{r}}$ can be obtained in an exact way using the \emph{arcsine law} ((cf. Eq. (36) in \cite{Mezghani-ISIT12}). Thus, the BLMMSE estimator $\hat{\mb{h}}^{\mc{Q}_1}_{\texttt{BLMMSE}}$ can be found in closed-form (cf. Eq. (14) in \cite{Li-Swindlehurst-TSP17}). However, when $b>1$, the covariance matrix $\mb{C}_{\mb{r}}$ cannot be analytically obtained in an exact way \cite{Mezghani-ISIT12}. Instead, an approximation of $\mb{C}_{\mb{r}}$ (cf. Eq. (28) in \cite{Mezghani-ISIT12}) can be obtained as follows:
\vspace{-0.15cm}
\begin{eqnarray} 
\mb{C}_{\mb{r}} \approx (1-\eta_q) \left((1-\eta_q)\mb{C}_{\mb{y}} + \eta_q\textrm{diag}(\mb{C}_{\mb{y}})\right)
\end{eqnarray}
which depends on $\mb{C}_{\mb{y}}$ and $\mb{\Phi}$ as a result.  
Thus, optimizing $\mb{\Phi}$ to minimize the sum MSE $\tr\{\mb{C}_{\bs{\varepsilon}}\}$ can be a challenging task, even with the BLMMSE estimator $\hat{\mb{h}}^{\mc{Q}_b}_{\texttt{BLMMSE}}$. In recent work \cite{Li-Swindlehurst-TSP17,Mo-TSP-2017}, the training matrix  $\mb{\Phi}$ was set to be column-wise orthogonal. Then, for i.i.d. channel vector, e.g., $\mb{C}_{\mb{h}} = \mb{I}_{MK}$, and i.i.d. Gaussian noise vector, e.g., $\mb{C}_{\mb{n}} = N_0\mb{I}_{M\tau}$, the diagonal elements of $\mb{C}_{\mb{y}}$ are $K\rho + N_0$. Thus,
\vspace{-0.15cm}
\begin{eqnarray}
\mb{C}_{\mb{r}} \approx (1-\eta_q) \left[(1-\eta_q)\bar{\mb{\Phi}}\bar{\mb{\Phi}}^H + (K\rho\eta_q + N_0) \mb{I}\right].
\end{eqnarray}
The BLMMSE estimator in this case can be simplified as 
\vspace{-0.15cm}
\begin{eqnarray}\label{BLMMSE}
	\hat{\mb{h}}^{\mc{Q}_b}_{\texttt{BLMMSE}} &=& \bar{\mb{\Phi}}^H\left((1-\eta_q)\bar{\mb{\Phi}}\bar{\mb{\Phi}}^H + (K\rho\eta_q + N_0)\,\mb{I}_{M\tau}\right)^{-1}\mb{r} \nonumber \\
	&=& 
	\left((1-\eta_q)\bar{\mb{\Phi}}^H\bar{\mb{\Phi}} + (K\rho\eta_q + N_0)\,\mb{I}_{M\tau}\right)^{-1}\bar{\mb{\Phi}}^H\mb{r} \nonumber \\
	&=& \frac{\bar{\mb{\Phi}}^H\mb{r}}{\rho\tau + N_0 + \rho\eta_q(K-\tau)}.
\end{eqnarray}

We will use the BLMMSE estimator $\hat{\mb{h}}^{\mc{Q}_1}_{\texttt{BLMMSE}}$ in \cite{Li-Swindlehurst-TSP17} and the above $\hat{\mb{h}}^{\mc{Q}_b}_{\texttt{BLMMSE}}$ estimator (for $b>1$) for benchmarking. The BLMMSE estimator is also applicable for estimating non-Gaussian channel vector $\mb{h}$ as long as $\mb{C}_{\mb{h}} = \mb{I}_{MK}$. We stress that the BLMMSE estimator is not optimal for MIMO systems with few-bit ADCs. While using $K$ columns of a $\tau\times \tau$ DFT matrix as the training signal in \cite{Li-Swindlehurst-TSP17} simplifies the BLMMSE estimator expression \eqref{BLMMSE}, it is easy to verify with numerical simulations that different combinations of the $K$ columns provide different sum MSE $\tr\{\mb{C}_{\mb{\varepsilon}}\}$ results. Moreover, no particular combination works best for the whole range of SNR or quantizing bit numbers. An exhaustive search over $\binom{\tau}{K}$ combinations can be prohibitively time consuming. 
This observation motivates us to study a DNN framework for optimizing the channel estimator and the training signal for few-bit MIMO systems.

\begin{table}
	\centering
	\caption{Optimum uniform quantizer for a Gaussian $\mc{N}(0,1)$ input \cite{Max-TIT-1960}.}
	\vspace{-.15cm}
	\begin{tabular}{|l|c|c|c|c|c|c|}
		\hline
		\textbf{Resolution} $b$ & $1$-bit & Ternary & $2$-bit & $3$-bit & $4$-bit \\
		\hline
		\textbf{Step-size} $\Delta_b$ & $\sqrt{8/\pi}$ & $1.224$ & $0.996$ & $0.586$ & $0.335$ \\
		\hline
		\textbf{Distortion} $\eta_b$ & $1-2/\pi$ & $0.1902$ & $0.1188$ & $0.0374$ & $0.0115$ \\
		\hline
	\end{tabular}
\vspace{-.25cm}
	\label{Table-1}
\end{table}

\begin{table}
	\centering
	\caption{Structure of the DNN regressor as an MMSE estimator.}
	\vspace{-.05cm}
	\begin{tabular}{|l|c|}
		\hline
		\textbf{Layer}& Output dimension \\
		\hline
		Input & $2\tau M$ \\
		Dense + ReLU & $2\tau M$ \\
		Dense + ReLU & $2\tau M$ \\
		Dense + Tanh & $2\tau M$ \\
		Dense & $2 KM$ \\
		\hline 
	\end{tabular}
\vspace{-.4cm}
	\label{Table-2}
\end{table}
\vspace{-.4cm}
\section{DNN-Optimized Channel Estimation and Training Signal Design}
\subsection{DNN-Optimized Channel Estimation}\label{MMSE-sect}
When an optimal MMSE estimator cannot be obtained analytically, a data-driven approach based on DNN can be used to approximate that estimator \cite{EE-278}. Given quantized observation $\mb{r}=\mc{Q}_b(\mb{y})$ and the channel model, a large data set on $(\mb{r},\mb{h})$ can be generated to train the DNN channel estimator. Since DNN can only work with real inputs and real outputs, 
we set the input as $\mb{r}^{\Re} = [\mr{Re}\{\mb{r}^T\},\mr{Im}\{\mb{r}^T\}]^T \in \mathbb{R}^{2\tau M}$ and the output as $\bar{\mb{h}}^{\Re} = [\mr{Re}\{\bar{\mb{h}}^T\},\mr{Im}\{\bar{\mb{h}}^T\}]^T \in \mathbb{R}^{2K M}$. 
We then consider a feed-forward DNN regressor with $3$ hidden layers, whose details are given in Table \ref{Table-2}. We use the Rectified Linear Unit (ReLU) activation function at the first two hidden layers and the $\mr{Tanh}$ activation function at the last one. We note that a similar DNN-based channel estimator with only fully connected layers has been proposed in \cite{Gao-Li-CLetters-2019} for the case of $\tau=1$. In this study, we also include the batch normalization technique for stabilizing the training process \cite{Batch_Norm}. A residual network \cite{ResNet} is also implemented by feeding the training data into the input of the last hidden layer. The cost function is set to minimize $\big\|\bar{\mb{h}}^{\Re} - \mb{h}^{\Re}\big\|^2$, where ${\mb{h}}^{\Re} = [\mr{Re}\{{\mb{h}}^T\},\mr{Im}\{{\mb{h}}^T\}]^T$. Effectively, the DNN regressor can be interpreted as a nonlinear MMSE channel estimator. 
We found that the DNN regressor presented in Table \ref{Table-2} performs very well for the purpose of optimizing the training sequence for few-bit MIMO systems, as presented in the next section. The DNN structure in Table \ref{Table-2} also provides a \emph{good} balance between complexity and performance, while avoiding the over-fitting issue. Certainly, it is interesting for future work to search for other DNN channel estimators with different layer types and activation functions. 

\begin{figure*}[htb!]
	\centering
	\includegraphics[width=135mm]{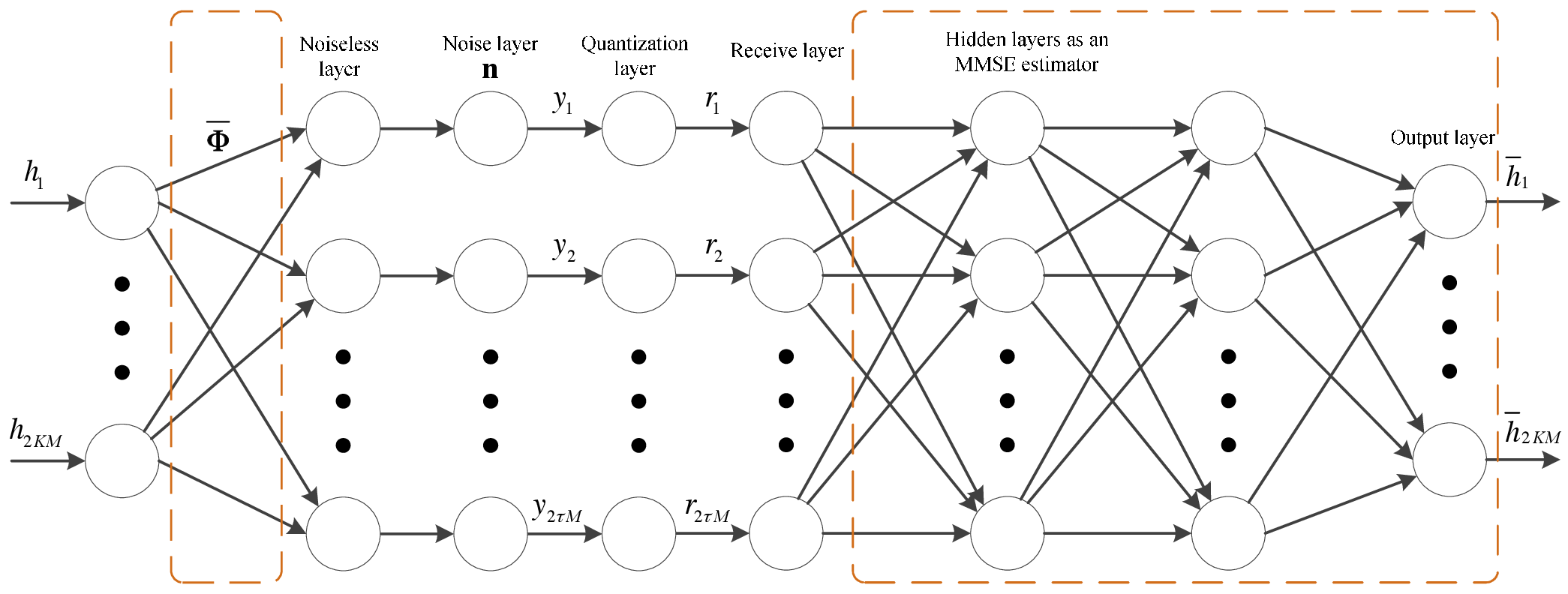}
	\vspace{-0.1cm}
	\caption{Diagram of an autoencoder to jointly optimize $\bs{\Phi}$ and the nonlinear MMSE estimator. The trainable parameters are enclosed in dashed boxes.}
	\label{Autoencoder}
\end{figure*}


\vspace{-0.25cm}
\subsection{Optimized Training Matrix Design using DNN}
Built upon the DNN regressor for channel estimation, we propose an autoencoder to optimize the training signal $\mb{\Phi}$.  Since DNN can only work with real numbers, we set $\mb{\Phi}^{\Re} = [\mr{Re}\{\mb{\Phi}^T\}, \mr{Im}\{\mb{\Phi}^T\}]^T \in \mathbb{R}^{2\tau \times K}$ as the variable to be optimized. Under a sum-power constraint, we can set a limit on the Frobenius norm of $\mb{\Phi}^{\Re}$. Likewise, a limit on the norm of each column of $\mb{\Phi}^{\Re}$ can be set to impose a per-antenna or per-user power constraint. Illustrated in Fig. \ref{Autoencoder} is the autoencoder, where it takes the channel vector $\mb{h}^{\Re}$ as input and reconstruct an estimated vector $\bar{\mb{h}}^{\Re}$ as output. While the second part of the autoencoder (from the ``Receive layer'') resembles the DNN channel estimator given in Table \ref{Table-2}, the novelty of the proposed autoencoder lies in its first part. Emulating the unquantized system model in vectorized form \eqref{vector-model} and the quantizer $\mb{r} = \mc{Q}_b(\mb{y})$, the operation of the autoencoder's first few layers are as follows:
\begin{itemize}
\item The ``Noiseless layer'' is used to obtain the complex multiplication $\bar{\mb{\Phi}}\mb{h}$. We split $\mb{\Phi}^{\Re}$ into two matrices $\mr{Re}\{\mb{\Phi}\}$ and $\mr{Im}\{\mb{\Phi}\}$ with the same size $\tau \times K$ 
and form matrix $\mb{\Phi}^{\Re\Im} = \left[\begin{array}{lr}  \mr{Re}\{\mb{\Phi}\} &-\mr{Im}\{\mb{\Phi}\} \\ \mr{Im}\{\mb{\Phi}\} & \mr{Re}\{\mb{\Phi}\}\end{array}\right]$. We then take the matrix multiplication of $\big(\sqrt{\rho}\,\mb{\Phi}^{\Re\Im} \otimes \mb{I}_M\big) \mb{h}^{\Re}$ at this layer to get a length-$2\tau M$ real-valued vector representing $\bar{\mb{\Phi}}\mb{h}$.
\item The ``Noise layer'' is used to generate the noise vector $\mb{n}^{\Re}$, which is added to the ``Noiseless layer'' to obtain a length-$2\tau M$ real-valued vector $\mb{y}^{\Re}$ representing $\mb{y}$ in \eqref{vector-model}.
\item The ``Quantization layer'' performs element-wise quantization on  $\mb{y}^{\Re}$  to obtain $\mb{r}^{\Re}$. For the $1$-bit quantizer, we use the $\mr{sign}$ function and $\mb{r}^{\Re}$ is comprised of $\pm 1$. For other quantizing schemes, we rely on the domain knowledge of $\mb{y}$ and use the optimal uniform quantizer for a Gaussian source presented in Section \ref{quantized-sect}. The standard deviation at $\sqrt{(K\rho + N_0)/2}$ per real/imaginary dimension is passed to this layer and used to scale the decision thresholds for quantizing. For the ternary quantizer, $\mb{r}^{\Re}$ is comprised of $\{-1,0,1\}$. Finally, for a $b$-bit quantizer, $\mb{r}^{\Re}$ is comprised of $\{\pm 1,\pm 3,\ldots, \pm 2^{b-1}-1\}$.
\item The ``Receive layer'' provides the input for the MMSE estimator, whose structure is explained in Section \ref{MMSE-sect}.
\end{itemize}

We note that the ``Noiseless layer''  is neither a convolutional layer nor a  fully connected dense layer in existing deep learning literature. More specifically, this layer enables the multiplications of $\mb{\Phi}^{\Re\Im}$ with partitions of $\mb{h}^{\Re}$, where each partition represents a length-$K$ complex-valued channel vector from $K$ transmit antennas to a receive antenna. By realizing its structure, we facilitate embedding $\mb{\Phi}$ into the weight matrix of this ``Noiseless layer''. Thus, once being trained, the proposed autoencoder prompts a DNN-optimized joint design for the training signal $\mb{\Phi}$ and the nonlinear MMSE estimator. 

We also note that the derivative of the quantization function at the ``Quantization layer'' is zero almost everywhere, making it incompatible with back-propagation. We thus adopt the ``straight-through estimator'' method  \cite{Bengio-STE-2013} to circumvent this issue. In fact, this method has been popularized for training DNN for image classification with binarized weights and activations in \cite{Bengio-NIPS-2015,Bengio-BNN-2016}, ternarized weights in \cite{Zhu-Ternary-ICLR-2017}, and low-bit weights, activations, and gradients in DoReFa-Net \cite{DoReFa}.

\vspace{-0.3cm}
\section{Simulation Results}
\vspace{-0.1cm}
This section presents numerical results comparing two schemes: i.) DNN-optimized channel estimator and training signal design and ii.) BLMMSE channel estimator and DFT-based training signal, in terms of the MSE $\frac{\|\mb{h}-\hat{\mb{h}}\|^2}{MK}$. We consider two representative channel models: one with i.i.d. complex Gaussian coefficients corresponding to the Rayleigh fading channel and another one with LoS coefficients.\footnote{The Rician fading model can be represented as a linear combination of the two channel models under consideration. The numerical results can be extended to cover the Rician fading model as well.}	Except the case  $K=8$ and $\tau=64$ where a randomized combination was chosen, an exhaustive search at each SNR value was conducted to find the best combination of the $K$ DFT columns that attains the lowest MSE by the BLMMSE channel estimator.
\begin{figure}[t!]
	\centering
		\vspace{-0.75cm}
	\includegraphics[width=85.00mm]{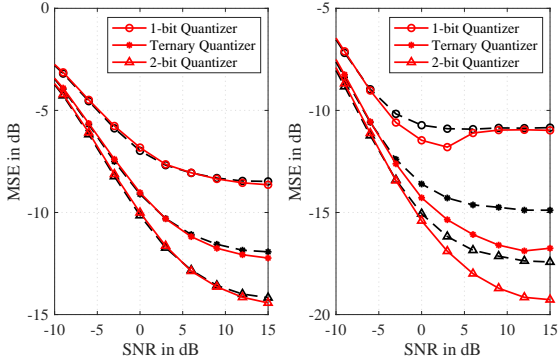}
	\caption{Left and right figures illustrate the MSE in estimating i.i.d. Gaussian channel coefficients with $\tau=16$ and $\tau=64$, respectively. The DNN channel estimator (results plotted in \emph{solid} lines) performs slightly better than the BLMMSE channel estimator (results plotted in \emph{dashed} lines) with less than $0.2$ dB in MSE reduction for $\tau=16$ and significantly  better for $\tau=64$.}
		\vspace{-0.5cm}
	\label{Figure-IID}
\end{figure}

In the first model, we assume $\mb{C}_{\mb{h}} = \mb{I}_{KM}$. We consider a multiuser scenario with $K=4$ and the training times of $\tau=16$ and $\tau=64$ symbols per user. When $\tau=16$, it is observed that the performances of the two schemes are comparable with negligible performance gains by the proposed scheme at high SNR. This result suggests the BLMMSE estimator with DFT-based training signal is an excellent option for low-bit systems with a short training duration. However, when $\tau=64$, the proposed DNN-based scheme significantly outperforms the BLMMSE scheme in ternary and $2$-bit systems. Intuitively, there are more degrees of freedom to optimize the training signal with a longer training duration. Interestingly, the proposed DNN-based scheme for $1$-bit system performs best at the SNR of $3$ dB.

The second model relies on LoS channel model studied in \cite{Abdelghany-2020} with $K=8$ users. We denote the angle-of-arrival  from user-$k$ as $\theta_k$ (with $-\pi/3\leq \theta_k\leq \pi/3$), corresponding to the spatial frequency $\Omega_k = 2\pi \frac{d}{\theta}\sin\theta_k$, where $\lambda$ is the carrier wavelength and $d$ set at half-wavelength is the inter-element spacing. The length-$M$ channel vector for user-$k$ is given by 
$\mb{h}_k = A_k e^{j\phi_k} \,\big[1,e^{j\Omega_k},e^{j2\Omega_k},\ldots,e^{j(M-1)\Omega_k}\big]^T$,
where $\phi_k$ is an arbitrary phase shift and $A_k$ depends on user-$k$'s location. Here, we randomly generate the user locations so that $\mathbb{E}[A_k^2] = 1$. It is easy to verify that the channel coefficients in $\mb{h}_k$ are zero-mean with unit variance and independent of each other, which makes $\mb{C}_{\mb{h}} = \mb{I}_{MK}$.  Note that ${\mb{h}}_k$ is \emph{not} Gaussian. However, by virtue of the central limit theorem, the noiseless received signal at an arbitrary antenna is well modeled as zero-mean complex Gaussian even for a moderate number of users (e.g., $K=8$) \cite{Abdelghany-2020}. Moreover, the noisy received signal at the quantizer input is also Gaussian with variance $K\rho + N_0$. Thus, we adopt the optimal uniform quantizer in Table \ref{Table-1} for both DNN-based and BLMMSE schemes. The analysis on BLMMSE channel estimator presented in Section \ref{quantized-sect} also stands, since $\mb{C}_{\mb{h}} = \mb{I}_{MK}$. As both $\mb{h}$ and $\mb{r}$ are not Gaussian, a linear MMSE estimator is far from the optimal MMSE estimator. Fig. 3 confirms the superior performance by the DNN-based scheme in both cases $\tau=16$ and $\tau=64$. This result indicates the potential of using DNN-based training signal designs for non-Gaussian channel estimation.

\begin{figure}[t!]
	\centering
		\vspace{-0.75cm}
	\includegraphics[width=85.00mm]{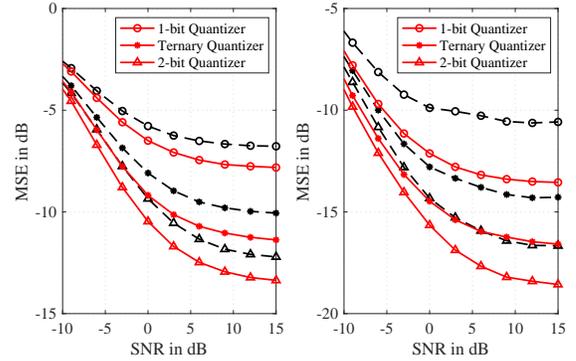}
	\caption{Left and right figures illustrate the MSE in estimating i.i.d. LoS channel coefficients with $\tau=16$ and $\tau=64$, respectively. The DNN channel estimator (results plotted in \emph{solid} lines) clearly outperforms the BLMMSE channel estimator (results plotted in \emph{dashed} lines) in both cases. Up to $3$ dB reduction of the MSE floors is observed by the proposed DNN-based scheme.}
		\vspace{-0.5cm}
	\label{Figure-LoS}
\end{figure}

	\vspace{-0.25cm}
\section{Conclusion}
This paper presents a DNN-based approach for estimating channel and designing training signals for MIMO systems with few-bit ADCs. It has shown an autoencoder structure whose first layer's weight matrix were designed to embed the training signal. Numerical results have demonstrated much lower MSE floors by the proposed DNN-based scheme than that by the Bussgang-based linear MMSE channel estimator with DFT-based training signals.
\balance
%

\end{document}